\documentclass[11pt]{article}
\usepackage[margin=1in]{geometry}
\usepackage{jheppub} 
\usepackage{amsmath,amssymb,mathtools,bm}
\usepackage{physics}
\usepackage{tikz-cd}
\usepackage{hyperref}
\usepackage{microtype}
\usepackage{enumitem}
\usepackage{bbm} 
\usepackage{xcolor} 
\usepackage{float}

\title{Hofstadter's Butterfly in AdS$_3$ Black Holes}
\author[a,b]{Kazuki Ikeda}
\emailAdd{kazuki.ikeda@umb.edu}
\affiliation[a]{Department of Physics, University of Massachusetts, Boston, MA 02125, USA}
\affiliation[b]{Center for Nuclear Theory, Department of Physics and Astronomy, Stony Brook University, Stony Brook, New York 11794-3800, USA}
\author[c]{Yaron Oz}
\affiliation[c]{School of Physics and Astronomy, Tel-Aviv University,
Tel-Aviv 69978, Israel}
\emailAdd{yaronoz@tauex.tau.ac.il}
\date{\today}

\abstract{
We derive the reduced Dirac Hamiltonian on the non-rotating BTZ background and use its redshift structure to construct a gauge-covariant single-band lattice model on the constant-time BTZ cylinder. In equal-area coordinates the AdS radius $L$ fixes the local Gaussian curvature, while the horizon radius $r_h$ fixes the throat size and the strength of the near-horizon redshift. The lattice model therefore has a direct geometric interpretation and is not presented as an unshown reduction of the two-component Dirac lattice. Its angular Fourier transform yields an exact curved Harper equation with BTZ-dependent hopping amplitudes and a consistent dimensionless angular quasi-momentum. We then supplement global parameter scans with state-resolved diagnostics: spectra color-coded by mean radius, local density of states, direct flux-response versus radius correlations, and Aharonov--Bohm spectral flow and persistent current on the BTZ cycle. These results show that weaker curvature sharpens the butterfly-like fragmentation, whereas larger horizons suppress both magnetic and Aharonov--Bohm response by creating weakly dispersing near-horizon states.
}
\begin{document}
\maketitle
\flushbottom

\section{Introduction}

Black holes continue to serve as a meeting ground for quantum gravity, thermal physics, and quantum many-body theory. The thermodynamic interpretation of horizon area and Hawking radiation established that gravity admits an intrinsically statistical description \cite{Bekenstein1973,Hawking1975}. In asymptotically anti-de Sitter (AdS) spacetimes this perspective becomes especially concrete. Three-dimensional gravity with negative cosmological constant admits the Bañados--Teitelboim--Zanelli (BTZ) black hole \cite{BTZ1992,BHTZ1993}, whose asymptotic symmetry algebra already anticipates a dual conformal field theory through the Brown--Henneaux central charge \cite{BrownHenneaux1986}. Along with the microscopic entropy counting of the BTZ black hole \cite{Strominger1998}, these results make BTZ one of the cleanest laboratories for AdS/CFT \cite{Maldacena1998,Aharony:1999ti,GKP1998,Witten1998}. The same holographic framework has also become a standard source of intuition for strongly correlated quantum matter, transport, and finite-density physics \cite{Hartnoll2009,McGreevy2010,Sachdev2012}.

A complementary lesson of the last two decades is that geometry in AdS is tightly tied to entanglement and information processing. Holographic entanglement entropy and its covariant extension tie areas to boundary entanglement data \cite{RyuTakayanagi2006,HRT2007}, while tensor-network and ``entanglement builds geometry'' viewpoints have emphasized the special role of negatively curved spatial graphs \cite{VanRaamsdonk2010,Swingle2012}. Black holes are also central in quantum-information diagnostics of scrambling \cite{HaydenPreskill2007}, and holography has sharpened the interpretation of bulk locality in terms of quantum error correction \cite{AlmheiriDongHarlow2015,Pastawski2015}. In parallel, hyperbolic tessellations have become concrete resources for quantum memories and low-overhead code constructions \cite{BreuckmannTerhal2016,BreuckmannEtAl2017}. These developments suggest that negative curvature is not merely a geometric decoration: it organizes robust structures across gravity, entanglement, and quantum computation.

From the condensed-matter side, magnetic lattice problems have long provided a canonical arena in which geometry, topology, and spectroscopy interact. The Harper and Hofstadter equations, along with the TKNN interpretation of quantized Hall response, established the modern language of magnetic bands and topological integers \cite{Harper1955,Hofstadter1976,TKNN1982}. Curved spaces enrich this problem already at the continuum level: charged particles on the hyperbolic plane exhibit Landau quantization and spectral structures that differ qualitatively from their Euclidean counterparts \cite{Comtet1987,CamporesiHiguchi1994}. More recently, hyperbolic band theory has generalized Bloch ideas to negatively curved lattices, leading to automorphic Bloch theorems, crystallographic classifications, high-dimensional irreducible representations, conformal field theory, and genuinely non-Abelian band phenomena \cite{MaciejkoRayan2021,MaciejkoRayan2022,KienzleRayan2022,BoettcherEtAl2022,ChengEtAl2022,UrwylerEtAl2022,Ikeda_2021,doi:10.1139/cjp-2022-0145,StegmaierEtAl2022,LenggenhagerEtAl2023,TummuruEtAl2024}. On the experimental side, synthetic platforms have begun to realize negatively curved graphs and their band topology in circuit-QED, topolectrical, and circuit-network architectures \cite{KollarEtAl2019,ZhangEtAl2022,LenggenhagerEtAl2022,ChenEtAl2023}. Very recent work has even started to simulate holographic conformal dynamics directly on hyperbolic lattices \cite{DeyEtAl2024}.

These threads motivate the central question of the present paper: what is the natural magnetic band problem associated with an AdS black hole? 
We identify a horizon-induced mechanism that generates a weakly dispersing, flux-insensitive sector absent in flat or purely hyperbolic systems.
The constant-time slice of the non-rotating BTZ geometry is locally hyperbolic, but it is not simply the hyperbolic plane. It also contains a horizon, a throat, and a non-contractible angular cycle. This makes BTZ a particularly sharp bridge between two lines of thought that are often discussed separately: hyperbolic band theory, which emphasizes negative curvature and non-Euclidean crystallography, and black-hole holography, which emphasizes redshift, spin structure, and cycle-sensitive observables. In such a geometry, one can ask not only how curvature deforms a Harper-Hofstadter spectrum, but also how near-horizon redshift suppresses angular motion, how Ramond and Neveu--Schwarz sectors enter an effective lattice problem, and how Aharonov--Bohm response behaves on the BTZ cycle.

In this work we adopt this perspective. We first derive the reduced Dirac Hamiltonian on the non-rotating BTZ background and identify the redshifted inverse proper bond lengths that control the effective single-band problem. We then formulate a gauge-covariant lattice model on the constant-time BTZ cylinder in equal-area coordinates, where the AdS radius $L$ sets the local Gaussian curvature and the horizon radius $r_h$ sets the throat size and near-horizon redshift strength. Its angular Fourier transform yields an exact curved Harper equation with BTZ-dependent hopping amplitudes and a consistent dimensionless angular quasi-momentum. Numerically, we complement global parameter scans with state-resolved diagnostics --- spectra color-coded by mean radius, local density of states, direct flux-response versus radius correlations, and Aharonov-Bohm spectral flow and persistent current. A clear picture thus emerges: $L$ controls the strength of the butterfly's deformation from the flat Harper limit, while $r_h$ governs the degree to which near-zero modes become horizon-localized and weakly dispersing.

The purpose of the paper is therefore twofold. Conceptually, it offers a concrete interface between AdS black-hole geometry and hyperbolic band theory that can be read from the perspectives of string theory, holography, quantum information, condensed matter, quantum simulation, and mathematical physics. Technically, it isolates a minimal BTZ-derived lattice model whose relation to geometry is explicit at every step, rather than hidden inside an unshown discretization of the two-component Dirac problem.

The rest of the paper is organized as follows. Section~\ref{sec:AdS_Ham} describes our model. Section~\ref{sec:BTZ} introduces the equal-area coordinate, the probe magnetic field, and the gauge-covariant BTZ-derived lattice model, whose angular Fourier transform yields the exact curved Harper equation with the appropriate dimensionless quasi-momentum and spin-structure dependence. Section~\ref{sec:setup} specifies the numerical setup and the diagnostics used throughout, including horizon weight, local density of states, direct flux-response measures, and Aharonov--Bohm observables. Section~\ref{sec:results} presents the numerical results and shows how the AdS radius $L$ and the horizon radius $r_h$ play sharply distinct roles: $L$ controls the curvature-induced warping of the butterfly-like spectrum, while $r_h$ controls the buildup of weakly dispersing near-horizon states and the suppression of both magnetic and cycle response. Section~\ref{sec:conclusion} closes with the broader implications of this BTZ-derived curved Harper problem for hyperbolic band theory, holography, and quantum simulation.

\section{$AdS_3$ Black Hole Hamiltonian}\label{sec:AdS_Ham}

\subsection{Geometry}

We will work in three-dimensional curved space-time with metric $g_{\mu\nu}$, related to 
the flat Minkowski metric 
$\eta_{ab}=\mathrm{diag}(-,+,+)$ by the vielbein $e_\mu^a$ by:
\begin{equation}
g_{\mu\nu} = e_\mu^a e_\nu^b \eta_{ab} \ .
\label{VB}
\end{equation}

Consider the non-rotating BTZ metric (AdS radius $L$, horizon $r_h$):
\begin{equation}
  ds^2 = - f(r)\,dt^2 + \frac{dr^2}{f(r)} + r^2\,d\phi^2,
  \qquad f(r) = \frac{r^2 - r_h^2}{L^2},\qquad (t\in\mathbb R,\, r\!\in[r_h,\infty),\, \phi\!\sim\!\phi+2\pi).
  \label{eq:BTZmetric}
\end{equation}
An orthonormal triad $e^a{}_\mu$ ($a=0,1,2$) is given by the 1-forms
\begin{equation}
  e^{0} = \sqrt{f}\,dt,\qquad
  e^{1} = \frac{dr}{\sqrt{f}},\qquad
  e^{2} = r\,d\phi \ .
\end{equation}
The torsionless spin connection $\omega^a{}_b$ obeys
Cartan's equation 
\begin{equation}
 de^a + \omega^a{}_b\wedge e^b=0 \ .    
\end{equation}
 This yields the nonzero 1-forms:
\begin{equation}
  \omega^{01} = \frac{f'(r)}{2}\,dt,\qquad
  \omega^{12} = -\sqrt{f(r)}\,d\phi \ ,
  \label{eq:spinconn}
\end{equation}
with $f'(r)=\dv{f}{r}=\tfrac{2r}{L^2}$. The determinant is $\sqrt{-g}=r$. We will denote
\begin{equation}
  \alpha(r)\equiv \sqrt{f(r)}=\frac{\sqrt{r^2-r_h^2}}{L},
\end{equation}
which plays the role of the local redshift factor.



\subsection{Action and Hamiltonian}
Let $\psi$ be a two-component spinor in $2{+}1$~D (gamma matrices conventions as in the apppendix).
The curved-space Dirac action (including a chemical potential $\mu$ for later convenience) is
\begin{equation}
  S=\int dt\,dr\,d\phi\,\sqrt{-g}\;
  \bar\psi\Big(i\,\gamma^\mu D_\mu - m + \mu\,\gamma^\mu \delta_\mu^{~0}\Big)\psi,
  \qquad D_\mu=\partial_\mu - \frac{1}{2}\,\omega_{\mu}^{ab}\Sigma_{ab} \ .
\end{equation}
Because $\sqrt{-g}=r$, it is convenient to flatten the inner product by the local rescaling
\begin{equation}
  \chi(r,\phi,t) \;\equiv\; r^{1/2}\,f(r)^{-1/4}\,\psi(r,\phi,t),
  \label{eq:rescale}
\end{equation}
for which the canonical equal-time anticommutator becomes 
\begin{equation}
\{\chi(\mathbf{x}),\chi^\dagger(\mathbf{x}')\}=\delta(r-r')\delta(\phi-\phi') \ .    
\end{equation}
Exploiting axial symmetry, expand in angular modes. Since the coordinate
$\phi$ is non-contractible in the exterior, we are free to choose periodic or antiperiodic along $\phi$: periodic (Ramond)
\begin{equation}
\chi(\phi+2\pi) = \chi(\phi)
\quad\Rightarrow\quad
\chi(t,r,\phi)
= 
\sum_{\ell\in\mathbb Z} \chi_{\ell}(t,r)\, e^{i\ell  \phi} \ ,    
\end{equation}
or antiperiodic (Neveu–Schwarz)
\begin{equation}
\chi(\phi+2\pi) = -\chi(\phi)
\quad\Rightarrow\quad
\chi(t,r,\phi)
= 
\sum_{\ell\in\mathbb Z} \chi_{\ell}(t,r)\, e^{i (\ell+\tfrac12) \phi} \ .
\end{equation}

The Lagrangian density for each angular sector $\ell$ becomes:
\begin{equation}
  \mathcal L_\ell
  = \chi_\ell^\dagger\!\Big[
  i\partial_t
  + i\,f(r)\,\sigma_x\partial_r + \frac{i}{2}f'(r)\sigma_x
  + \,\frac{\sqrt{f(r)}}{r}\,\ell_{(R,NS)}\,\sigma_y
  - \sqrt{f(r)}\,m\,\sigma_z
  + \sqrt{f(r)}\,\mu
  \Big]\chi_\ell\; 
  \label{eq:Lell}
\end{equation}
where $\ell_{(R)}= \ell,~ \ell_{(NS)} = \ell+\frac{1}{2}$.

From \eqref{eq:Lell} we read the single-$\ell$ Hamiltonian \footnote{Note the structure of the hermitian operator $i\,A(r)\,\partial_r\;+\;\frac{i}{2}\,A'(r)$, where $A(r) = -f(r)\,\sigma_x$.}:
\begin{equation}
  H_\ell
  = \int_{r_h}^{\infty}\! dr\;\chi_\ell^\dagger(r)\,\Big[
    -\,i\,f(r)\,\sigma_x\partial_r -  \frac{i}{2}f'(r)\sigma_x - \,\frac{\sqrt{f(r)}}{r}\,\ell_{(R,NS)}\,\sigma_y
      + \sqrt{f(r)}\,m\,\sigma_z
  - \sqrt{f(r)}\,\mu \Big] \chi_\ell(r)\; .
  \label{eq:Hell}
\end{equation}
The difference compared to the $AdS_2$ \cite{Ikeda:2025lig,Ikeda:2025yqq} is only the term $\frac{\sqrt{f(r)}}{r}\,\ell_{(R,NS)}\,\sigma_y$.

\section{BTZ-derived 2D lattice formulation}\label{sec:BTZ}

\subsection{Equal-area coordinate and spatial geometry}

A convenient radial variable is
\begin{equation}
 x\equiv\sqrt{r^2-r_h^2},
 \qquad
 r(x)=\sqrt{x^2+r_h^2},
 \qquad
 \alpha(x)=\frac{x}{L}.
 \label{eq:xcoord-strong}
\end{equation}
In these coordinates the constant-time spatial metric becomes
\begin{equation}
 ds_{\Sigma}^2=\frac{L^2}{r(x)^2}dx^2+r(x)^2 d\phi^2,
 \qquad
 \sqrt{g_{\Sigma}}=L,
 \label{eq:spatialmetric-strong}
\end{equation}
so every plaquette in an $(x,\phi)$ lattice of spacings $(a_x,a_{\phi})$ has the same area,
\begin{equation}
 A_p=L\,a_x a_{\phi}.
\end{equation}
The spatial spin connection is
\begin{equation}
 \omega^{12}=-\frac{x}{L}d\phi,
 \qquad
 d\omega^{12}=-\frac{1}{L}dx\wedge d\phi=-\frac{1}{L^2}e^1\wedge e^2,
\end{equation}
so the Gaussian curvature of the spatial slice is
\begin{equation}
 K=-\frac{1}{L^2}.
 \label{eq:gaussian-curvature-strong}
\end{equation}
This makes the roles of $L$ and $r_h$ transparent: $L$ controls the local negative curvature, whereas $r_h$ does \emph{not} change $K$ and instead fixes the minimal circumference $2\pi r_h$ and the strength of the near-horizon redshift.

The proper bond lengths of a rectangular equal-area lattice are
\begin{equation}
 \ell_x(x)=\frac{L}{r(x)}a_x,
 \qquad
 \ell_{\phi}(x)=r(x)a_{\phi}.
 \label{eq:proper-bonds-strong}
\end{equation}
Measured in boundary time, a local energy scale is redshifted by $\alpha(x)$, so the natural BTZ bond energies are
\begin{equation}
 t_x(x)=\frac{\alpha(x)}{\ell_x(x)}=\frac{x\,r(x)}{L^2 a_x},
 \qquad
 t_{\phi}(x)=\frac{\alpha(x)}{\ell_{\phi}(x)}=\frac{x}{L\,r(x)a_{\phi}}.
 \label{eq:redshifted-bonds-strong}
\end{equation}
These are precisely the two geometric coefficients that will appear in the effective lattice model.

Near the horizon,
\begin{equation}
 t_x(x)\sim \frac{r_h}{L^2 a_x}x,
 \qquad
 t_{\phi}(x)\sim \frac{1}{Lr_h a_{\phi}}x,
 \qquad x\to 0,
 \label{eq:near-horizon-scaling-strong}
\end{equation}
so the angular magnetic coupling vanishes linearly at the horizon. 
We therefore expect states localized at small \(x\) to be weakly dispersing and only weakly sensitive to magnetic and Aharonov-Bohm flux.

\subsection{Probe magnetic field and the continuum BTZ Dirac operator}

We couple the BTZ Dirac fermion to a probe $U(1)$ field,
\begin{equation}
 D_{\mu}\to D_{\mu}-iqA_{\mu}.
\end{equation}
A locally constant magnetic field on the BTZ spatial slice is obtained in the gauge
\begin{equation}
 A_x=0,
 \qquad
 A_{\phi}(x)=BLx+\frac{\Phi}{2\pi},
 \label{eq:Landau-BTZ-strong}
\end{equation}
where $B$ is the local magnetic field and $\Phi$ is an Aharonov--Bohm flux through the non-contractible $\phi$-cycle. Indeed,
\begin{equation}
 F_{x\phi}=\partial_x A_{\phi}=BL,
 \qquad
 F_{\hat x\hat \phi}=B.
\end{equation}
After the additional rescaling
\begin{equation}
 \eta(x,\phi,t)=\sqrt{\frac{x}{r(x)}}\,\chi(r(x),\phi,t),
 \label{eq:eta-strong}
\end{equation}
the two-dimensional single-particle BTZ Dirac Hamiltonian becomes
\begin{equation}
 H_{\rm D}[A]=\int_0^{\infty}dx\int_0^{2\pi}d\phi\;\eta^{\dagger}
 \Big[
 -iv_x(x)\sigma_x\partial_x
 -\frac{i}{2}v_x'(x)\sigma_x
 +iv_{\phi}(x)\sigma_y\bigl(\partial_{\phi}-iqA_{\phi}(x)\bigr)
 +M(x)\sigma_z
 +V_{\rm loc}(x)
 \Big]\eta,
 \label{eq:H2D-dirac-strong}
\end{equation}
with
\begin{equation}
 v_x(x)=\frac{x\,r(x)}{L^2},
 \qquad
 v_{\phi}(x)=\frac{x}{L\,r(x)},
 \qquad
 M(x)=\frac{x}{L}m.
 \label{eq:vx-vphi-strong}
\end{equation}
Here $V_{\rm loc}(x)$ denotes an optional redshifted scalar potential. In the numerical study below we set $m=0$ and $V_{\rm loc}=0$.

Equation \eqref{eq:H2D-dirac-strong} shows precisely how BTZ geometry enters the two-dimensional problem: through the directional coefficients $v_x$ and $v_{\phi}$, along with the equal-area plaquette flux implied by \eqref{eq:Landau-BTZ-strong}. A direct two-component lattice discretization of \eqref{eq:H2D-dirac-strong} would require an explicit treatment of fermion doubling, for example with a Wilson or staggered regularization. In this paper we instead study the one-band effective model defined next.

\subsection{Effective BTZ Harper model}

Let $c_{n,j}$ be a single-band degree of freedom on the equal-area lattice,
\begin{equation}
 x_n=x_0+\left(n+\frac12\right)a_x,
 \qquad
 \phi_j=ja_{\phi},
 \qquad
 a_{\phi}=\frac{2\pi}{N_{\phi}},
 \qquad
 n=0,\dots,N_x-1,
 \quad
 j=0,\dots,N_{\phi}-1.
 \label{eq:lattice-grid-strong}
\end{equation}
We define the gauge-covariant nearest-neighbour Hamiltonian
\begin{equation}
 H_{\rm eff}=
 -\sum_{n,j}\Big[t_{n+\frac12}\,c^{\dagger}_{n+1,j}c_{n,j}+\lambda_n e^{-iq a_{\phi}A_{\phi}(x_n)}c^{\dagger}_{n,j+1}c_{n,j}+\text{h.c.}\Big]
 +\sum_{n,j}V_n c^{\dagger}_{n,j}c_{n,j},
 \label{eq:Heff2D-strong}
\end{equation}
with
\begin{equation}
 t_{n+\frac12}=\frac{x_{n+\frac12}r(x_{n+\frac12})}{L^2a_x},
 \qquad
 \lambda_n=\frac{x_n}{Lr(x_n)a_{\phi}}=\frac{N_{\phi}}{2\pi}\frac{x_n}{Lr(x_n)}.
 \label{eq:heff-coeffs-strong}
\end{equation}
This is the model solved numerically in the remainder of the paper. As written, it is a BTZ-cylinder generalization of the magnetic tight-binding problems introduced by Harper and Hofstadter and of their topological interpretation in the integer quantum Hall setting \cite{Harper1955,Hofstadter1976,TKNN1982}.

Note that while (\ref{eq:Heff2D-strong}) is a BTZ-derived effective lattice model, it is not presented as the exact spectrum of the continuum Dirac Hamiltonian \eqref{eq:H2D-dirac-strong}; rather, it is the minimal single-band implementation that retains the same BTZ redshift and magnetic geometry.
The relation to BTZ geometry is explicit. The radial and angular hopping amplitudes are the redshifted inverse proper bond lengths \eqref{eq:redshifted-bonds-strong}; the plaquette area is the BTZ equal-area cell $A_p=L a_x a_{\phi}$; and the Peierls phase is generated by the BTZ magnetic gauge field \eqref{eq:Landau-BTZ-strong}.
Thus the effective model retains the correct geometric scaling of energy and flux coupling but does not reproduce the full spinor structure or spectrum of the continuum Dirac operator.

Fourier transforming in the angular direction with a dimensionless quasi-momentum,
\begin{equation}
 c_{n,j}=\frac{1}{\sqrt{N_{\phi}}}\sum_{s=0}^{N_{\phi}-1}e^{i\kappa_s j}c_n(\kappa_s),
 \qquad
 \kappa_s=\frac{2\pi}{N_{\phi}}(s+\nu),
 \qquad \nu=0\ \text{or}\ \frac12,
 \label{eq:kphi-RNS-strong}
\end{equation}
we obtain an exact one-dimensional Harper equation,
\begin{equation}
 E\psi_n=
 -t_{n+\frac12}\psi_{n+1}
 -t_{n-\frac12}\psi_{n-1}
 +2\lambda_n\cos\!\Big(\frac{\kappa_{\phi}}{a_\phi}-2\pi\alpha_B\bigl(n+\tfrac12+\tfrac{x_0}{a_x}\bigr)-\frac{q\Phi}{N_{\phi}}\Big)\psi_n
 +V_n\psi_n,
 \label{eq:curved-harper-exact-strong}
\end{equation}
where
\begin{equation}
 \alpha_B\equiv\frac{qBLa_xa_{\phi}}{2\pi}.
 \label{eq:alphaB-strong}
\end{equation}
Equation \eqref{eq:curved-harper-exact-strong} is therefore an exact angular reduction of the effective BTZ lattice model \eqref{eq:Heff2D-strong}. 

\section{Numerical setup and diagnostics}\label{sec:setup}

Unless stated otherwise, spectra and observables are computed for the open radial chain \eqref{eq:curved-harper-exact-strong} with $V_n=0$. We use a common parameter block across the full analysis,
\begin{equation}
 N_x=N_{\phi}=48,
 \qquad
 a_x=0.10,
 \qquad
 \alpha_B\in[0,1],
 \qquad
 N_{\alpha}=241,
\end{equation}
with a scan over
\begin{equation}
 L\in\{8,10,12,14,16,18,20\},
 \qquad
 r_h\in\{0.3,0.6,1.0,1.5,2.0,3.0,4.0\}.
\end{equation}
Unless otherwise stated we set $\nu=0$ and $\Phi=0$.

The half-bandwidth is defined by
\begin{equation}
 E_{\rm bw}=\max |E|,
 \label{eq:halfbandwidth-strong}
\end{equation}
and the global flux sensitivity by
\begin{equation}
 S=\frac{1}{E_{\rm bw}}\left\langle \mathrm{std}_{\alpha_B}\,E_m^{\rm sort}(\alpha_B)\right\rangle_m,
 \label{eq:flux-sens-strong}
\end{equation}
where $E_m^{\rm sort}(\alpha_B)$ denotes the $m$th eigenvalue after sorting the full spectrum at fixed $\alpha_B$. The apparent box-counting dimension is extracted from
\begin{equation}
 N_{\rm box}(G)\sim G^{D_{\rm box}},
 \qquad
 G\in\{32,48,64,96,128\},
 \label{eq:box-dim-strong}
\end{equation}
using the point cloud $(\alpha_B,E/E_{\rm bw})$.

For each pair $(\alpha_B,\kappa_{\phi})$ we select the $n_{\rm mid}=4$ eigenstates with the smallest $|E|$. Their averages define the near-zero inverse participation ratio,
\begin{equation}
 \mathrm{IPR}=\sum_n |\psi_n|^4,
\end{equation}
and the second-moment length,
\begin{equation}
 \xi_2^2=\sum_n (x_n-\bar x)^2|\psi_n|^2,
 \qquad
 \bar x=\sum_n x_n|\psi_n|^2.
 \label{eq:xi2-strong}
\end{equation}
The near-zero density of states (DOS) is estimated from the window
\begin{equation}
 \rho_0=\frac{1}{\delta E}\left\langle \Theta\!\left(\frac{\delta E}{2}-|E|\right)\right\rangle,
 \qquad
 \delta E=0.05\,E_{\rm bw}.
 \label{eq:rho0-strong}
\end{equation}

To expose the spatial mechanism directly, we also use the mean radius
\begin{equation}
 \bar x_m=\sum_n x_n|\psi_n^{(m)}|^2,
\end{equation}
and the near-horizon weight
\begin{equation}
 W_h^{(m)}=\sum_{x_n<x_c}|\psi_n^{(m)}|^2,
 \qquad x_c=1.0.
 \label{eq:Wh-strong}
\end{equation}
The local density of states is
\begin{equation}
 \rho(x,E)=\left\langle\sum_m |\psi_m(x)|^2\,\delta_{\eta}(E-E_m)\right\rangle_{\alpha_B,\kappa_{\phi}},
 \qquad
 \eta=0.03\,E_{\rm bw},
 \label{eq:ldos-strong}
\end{equation}
where $\delta_{\eta}$ is implemented as a Gaussian broadening.

To quantify the state-resolved magnetic response at fixed $\alpha_B$, we define
\begin{equation}
 \mathcal F_m=\frac{|E_m(\alpha_*+\Delta\alpha)-E_m(\alpha_* - \Delta\alpha)|}{2\Delta\alpha\,E_{\rm bw}},
 \qquad
 \alpha_*=0.50,
 \qquad
 \Delta\alpha=0.01.
 \label{eq:state-flux-response-strong}
\end{equation}

Finally, we probe the non-contractible BTZ cycle by scanning the Aharonov--Bohm flux $\Phi$. At half filling we define the ground-state energy by filling the lowest $N_xN_{\phi}/2$ single-particle levels,
\begin{equation}
 E_{\rm GS}(\Phi)=\sum_{m=1}^{N_xN_{\phi}/2} E_m^{\rm sort}(\Phi),
\end{equation}
with persistent current
\begin{equation}
 I(\Phi)=-\pdv{E_{\rm GS}}{\Phi},
 \label{eq:current-strong}
\end{equation}
and an Aharonov--Bohm stiffness magnitude
\begin{equation}
 \mathcal D_{\Phi}=\frac{1}{E_{\rm bw}}\left|\pdv[2]{E_{\rm GS}}{\Phi}\right|_{\Phi=0}.
 \label{eq:stiffness-strong}
\end{equation}
Because the half-filled gap of a periodic radial closure is not uniform across the full parameter scan, we do not use Berry curvature as a primary bulk diagnostic here. The central claims are instead based on the open-chain spectra, the state-resolved radial observables, and the BTZ cycle response.

\section{Results}\label{sec:results}

\subsection{State-resolved spectra and horizon localization}

\begin{figure}[t]
\centering
\includegraphics[width=0.96\textwidth]{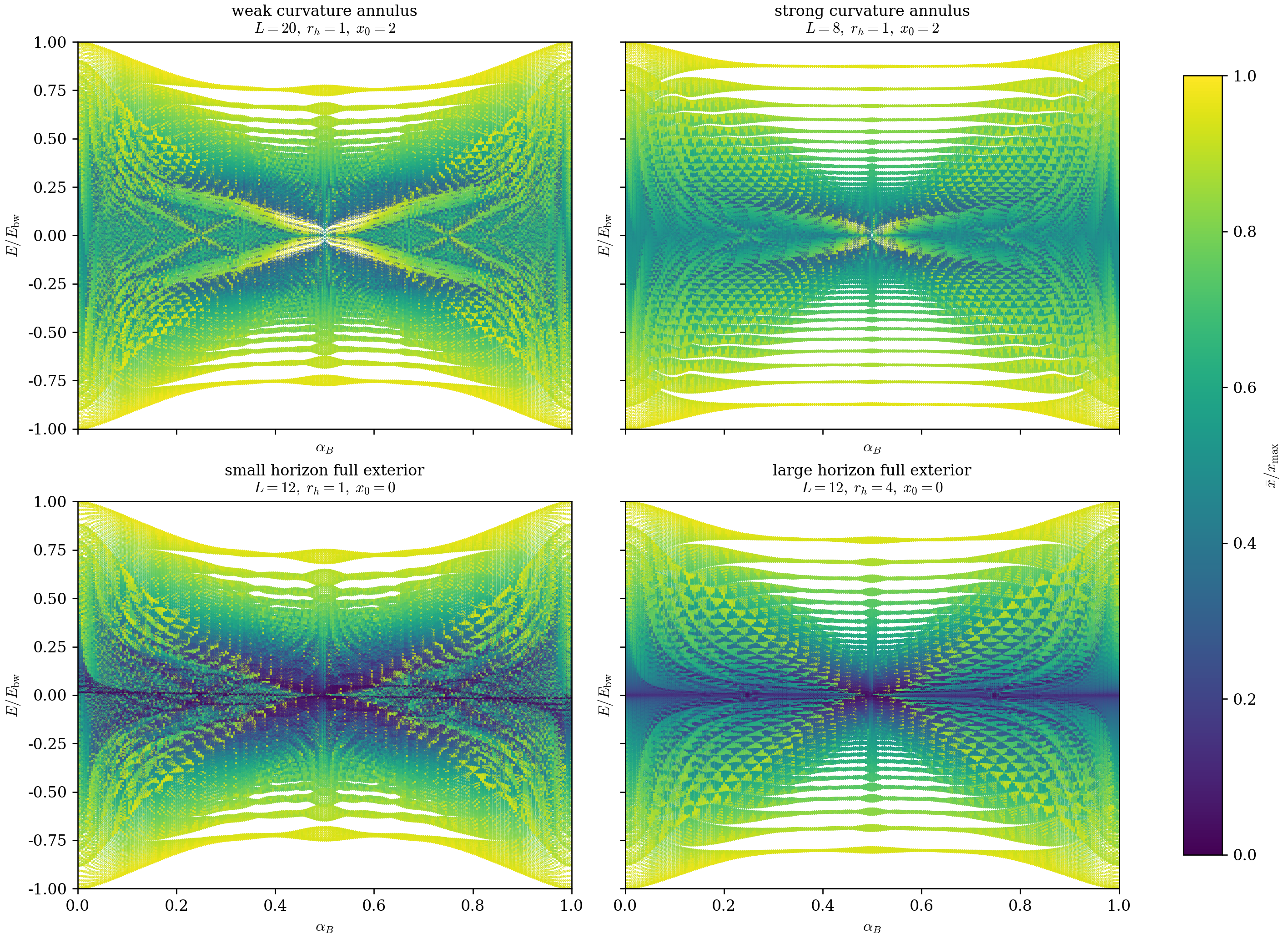}
\caption{Representative BTZ-derived fractal spectra color-coded by the normalized mean radius $\bar x/x_{\max}$. The annular panels isolate the curvature dependence through $K=-1/L^2$, while the full-exterior panels show that the nearly vertical central structures are the most horizon-localized states and therefore the least affected by the angular Harper modulation.}
\label{fig:coloredspectra-strong}
\end{figure}

We expect curvature to control the global deformation of the spectrum, while near-horizon redshift should produce weakly dispersing central branches.
Figure~\ref{fig:coloredspectra-strong} shows representative spectra of the exact BTZ Harper equation \eqref{eq:curved-harper-exact-strong}, with every point color-coded by the normalized mean radius $\bar x/x_{\max}$. This immediately separates the outer and inner parts of the spectrum. On an annulus, increasing $L$ sharpens the fragmented structure and moves the spectrum closer to the flat Harper pattern, while stronger curvature produces visibly more warping. Because $K=-1/L^2$, this annular comparison isolates a genuine curvature effect: larger $L$ weakens the local geometric warping without changing the BTZ cycle itself. On the full BTZ exterior, the weakly dispersing states near the center of the spectrum have the smallest $\bar x/x_{\max}$ and are therefore localized closest to the horizon. The large-$r_h$ panel makes this most transparent: the almost vertical central branches are not generic low-energy bands, but near-horizon branches. Their weak dispersion is precisely what one expects from the near-horizon scaling $t_{\phi}(x)\sim x/(Lr_h a_{\phi})$: increasing $r_h$ leaves the local curvature fixed, but makes angular motion increasingly ineffective for states that remain at small $x$.

\begin{figure}[t]
\centering
\includegraphics[width=0.96\textwidth]{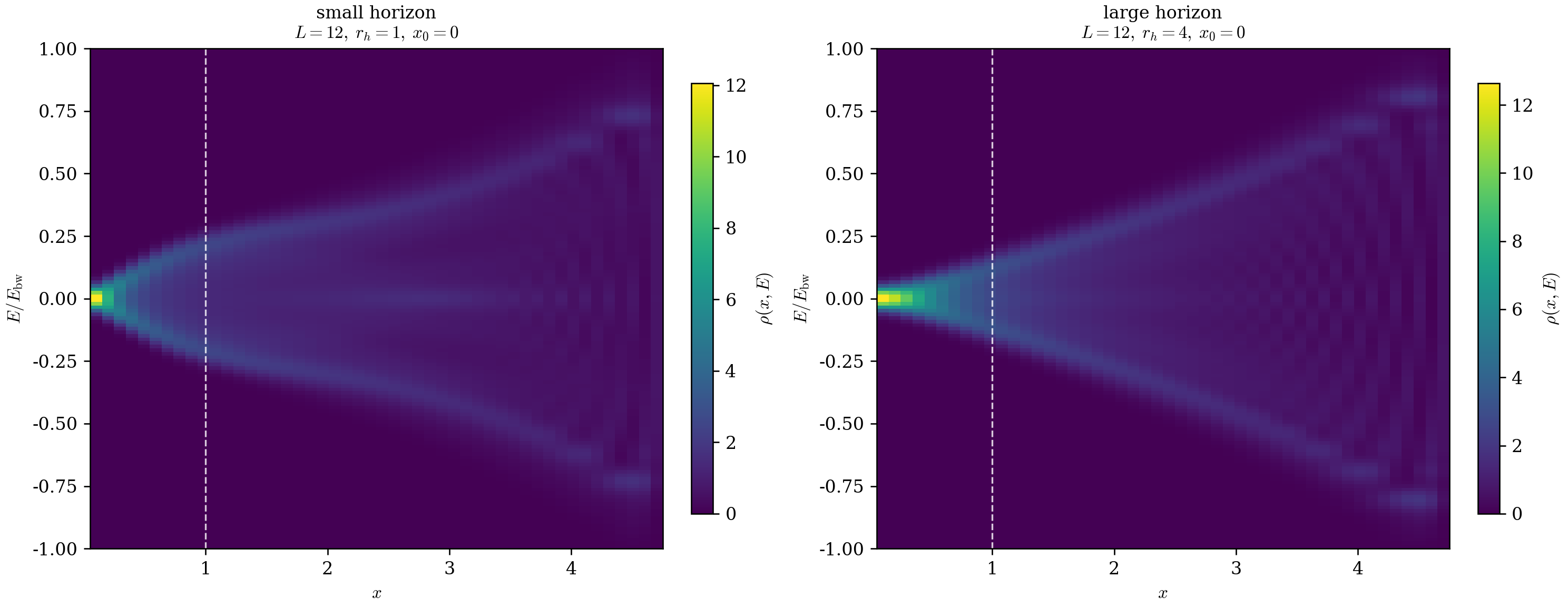}
\caption{Local density of states $\rho(x,E)$ on the full BTZ exterior. The dashed line marks the near-horizon window used in the definition of $W_h$. Larger $r_h$ transfers more near-zero spectral weight toward small $x$, which is precisely where the BTZ angular coupling vanishes linearly.}
\label{fig:ldos-strong}
\end{figure}

The same mechanism is visible in the local DOS. Figure~\ref{fig:ldos-strong} shows $\rho(x,E)$ on the full exterior at fixed $L=12$. The near-zero spectral weight is concentrated at small $x$, and this concentration becomes stronger as $r_h$ grows. This directly confirms that the low-energy peak is not merely a global redistribution of states: it is a horizon-localized contribution whose coupling to the magnetic modulation is reduced by the BTZ redshift. In other words, the central spectral smearing is produced by states that live where the angular hopping scale $t_{\phi}(x)$ is smallest. The comparison between the two panels also explains why $\rho_0$ can increase while global response measures decrease: a larger horizon packs more low-energy weight into the near-horizon region, but those extra states are dynamically less mobile around the BTZ circle and therefore contribute little to either magnetic or Aharonov--Bohm dispersion.

\subsection{Direct flux-response versus radius correlation}

\begin{figure}[t]
\centering
\includegraphics[width=0.96\textwidth]{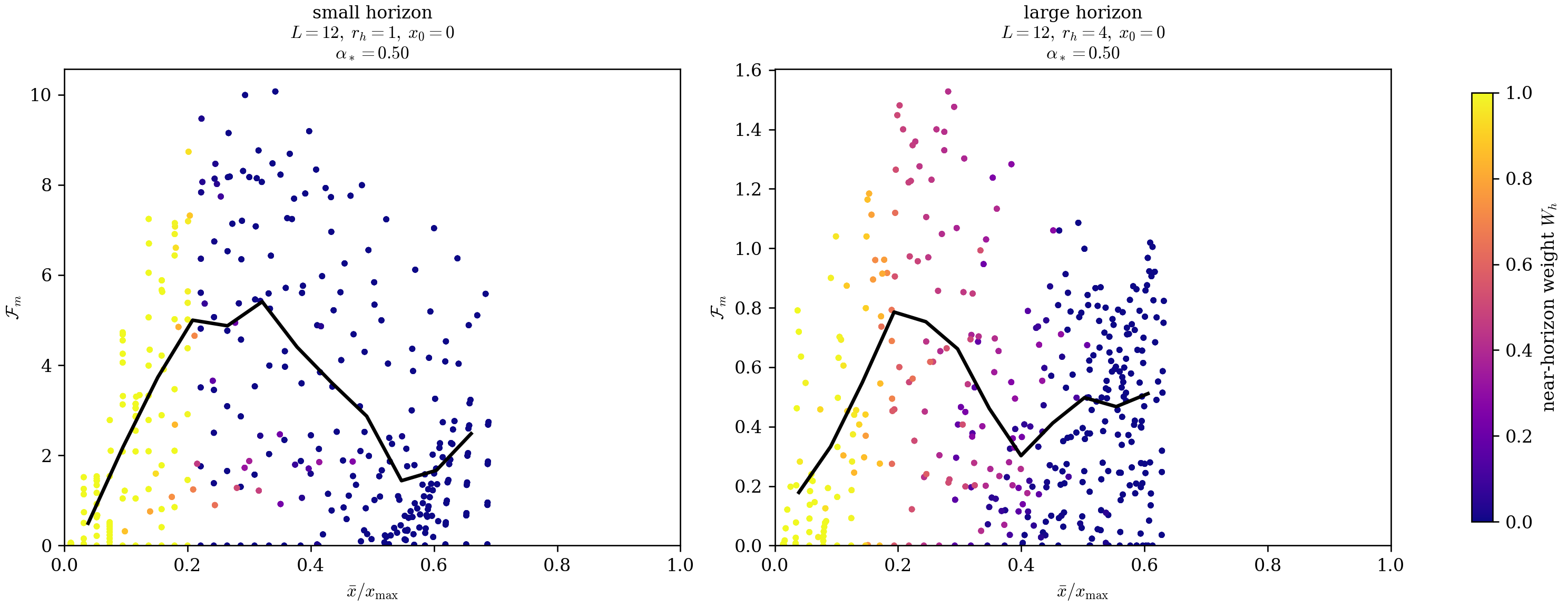}
\caption{State-resolved normalized flux response $\mathcal F_m$ versus the normalized mean radius $\bar x/x_{\max}$ at $\alpha_*=0.50$. Marker color encodes the near-horizon weight $W_h$, and the black curve shows the binned mean response. States closer to the horizon are systematically less flux-sensitive and carry the largest near-horizon weight.}
\label{fig:scatter-strong}
\end{figure}

The horizon mechanism can be tested more directly by looking at the state-resolved flux response \eqref{eq:state-flux-response-strong}. Figure~\ref{fig:scatter-strong} plots $\mathcal F_m$ against $\bar x/x_{\max}$ for representative full-exterior cases. The binned trend falls toward the horizon, showing that small-$\bar x$ states are systematically less responsive to magnetic flux, and
provides a direct confirmation of the near-horizon analytic scaling
$t_{\phi}(x)\sim x$, linking spatial localization to suppressed flux response.
As $x\to 0$, the BTZ angular hopping scale goes to zero, so states that remain close to the horizon acquire only a weak Harper modulation. The color scale adds an independent localization diagnostic: states with the largest near-horizon weight $W_h$ are precisely the ones clustered at small $\bar x$ and small $\mathcal F_m$. The large-$r_h$ case simply contains more such states, so the suppression of flux response is not an accidental low-energy feature but a direct consequence of radial localization near the throat.

\subsection{Aharonov--Bohm response on the BTZ cycle}

\begin{figure}[t]
\centering
\includegraphics[width=0.96\textwidth]{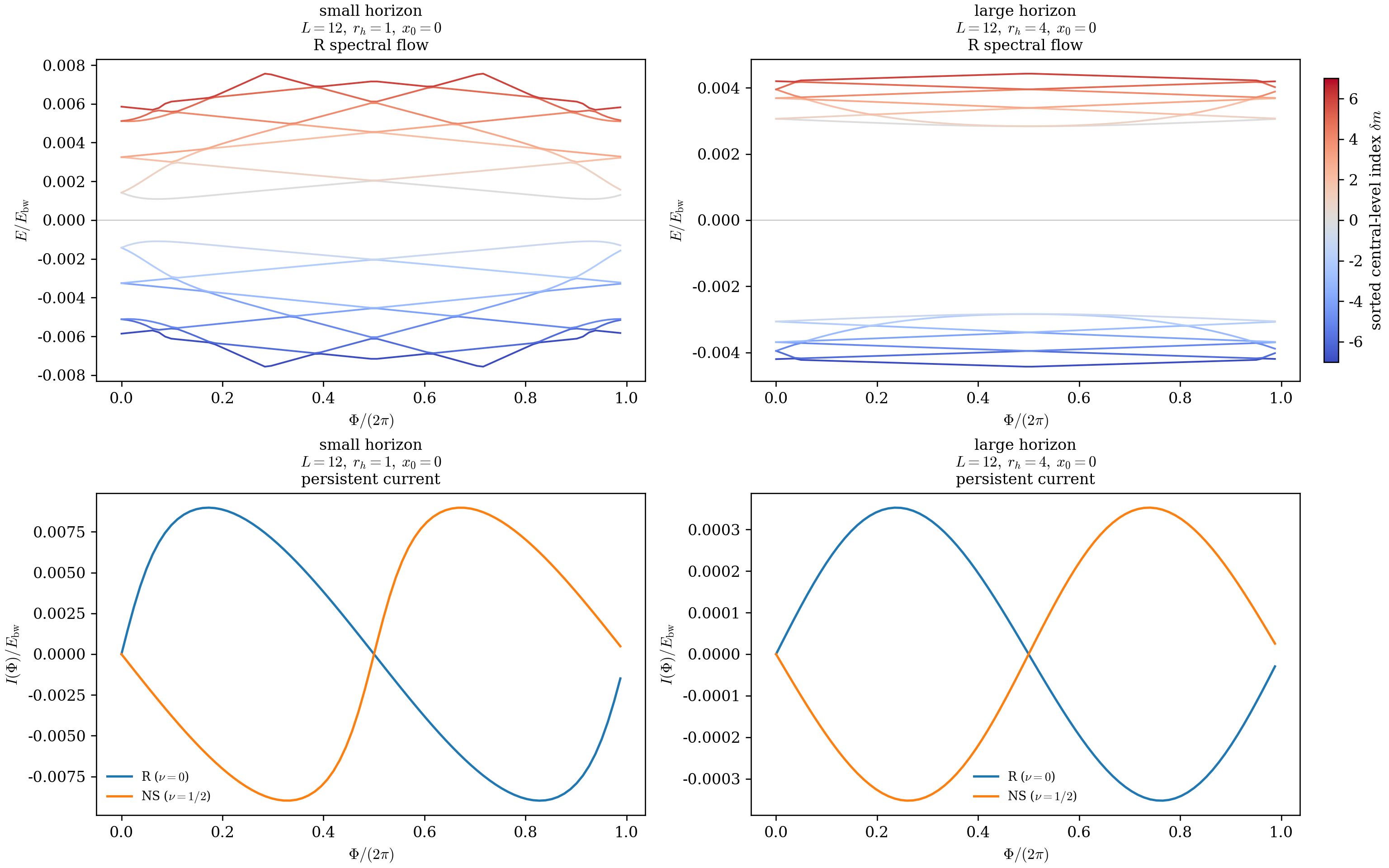}
\caption{Top: Aharonov--Bohm spectral flow of the central Ramond levels at $\alpha_B=0.50$. The shared colorbar labels the sorted central-level index $\delta m$ within the displayed window. Bottom: persistent current for Ramond and Neveu--Schwarz boundary conditions. Larger horizons strongly suppress the cycle response, while the R/NS shift is visible already at the level of the current.}
\label{fig:ab-strong}
\end{figure}

The BTZ exterior has a non-contractible $\phi$-cycle, so the Aharonov--Bohm flux $\Phi$ is a particularly natural probe. Figure~\ref{fig:ab-strong} shows the spectral flow of the central levels in the Ramond sector and the corresponding persistent current for both Ramond and Neveu--Schwarz boundary conditions. In the top row the branch color labels the sorted central-level index $\delta m$ within the displayed window, so the flattening at large $r_h$ can be seen as a collective narrowing of the whole central manifold rather than as a single accidental crossing. Two facts then stand out.

First, larger $r_h$ strongly flattens the spectral flow. This is the cycle-response version of the same horizon-redshift mechanism already seen in the $\alpha_B$ scan: when more low-energy states live at small $x$, the BTZ circle becomes less effective at dragging them. The bottom row shows the many-body consequence of the same effect: the persistent-current amplitude drops along with the spectral-flow width, which is exactly why the Aharonov--Bohm stiffness decreases in the heatmaps. Second, the Ramond and Neveu--Schwarz sectors are shifted with respect to one another, exactly as expected from the half-step shift in $\kappa_s$. This makes the spin structure a dynamical part of the effective model rather than a decorative boundary-condition choice.

\subsection{Global parameter scans and robustness}

\begin{figure}[t]
\centering
\includegraphics[width=\textwidth]{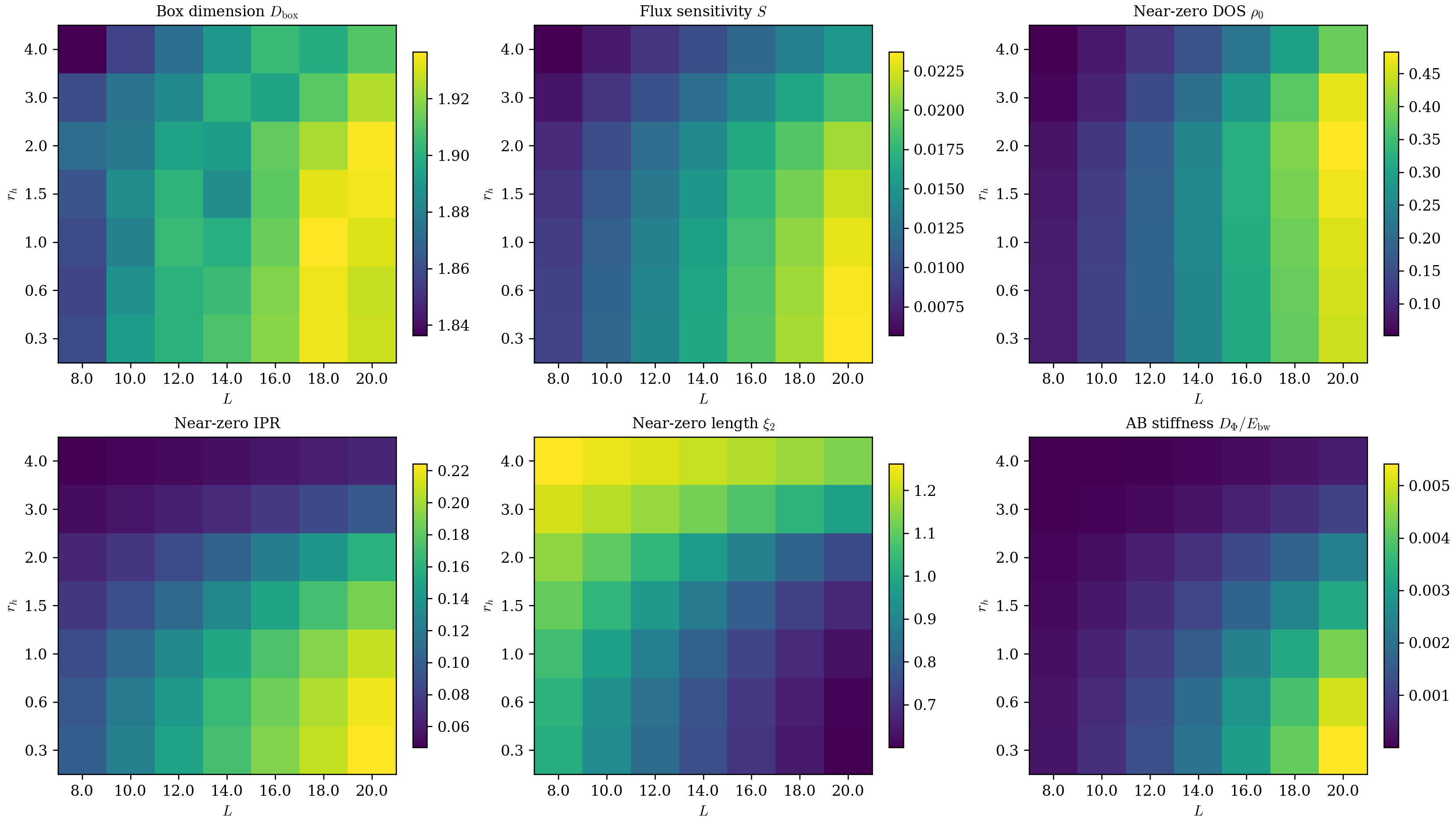}
\caption{Annular diagnostics of the BTZ-derived Harper model as functions of $L$ and $r_h$: the apparent box dimension $D_{\rm box}$, the global flux sensitivity $S$, the near-zero DOS $\rho_0$, the near-zero IPR, the second-moment length $\xi_2$, and the Aharonov--Bohm stiffness magnitude $\mathcal D_{\Phi}$. Larger $L$ sharpens the fragmented structure and enhances the cycle response.}
\label{fig:annulusmetrics-strong}
\end{figure}

\begin{figure}[t]
\centering
\includegraphics[width=\textwidth]{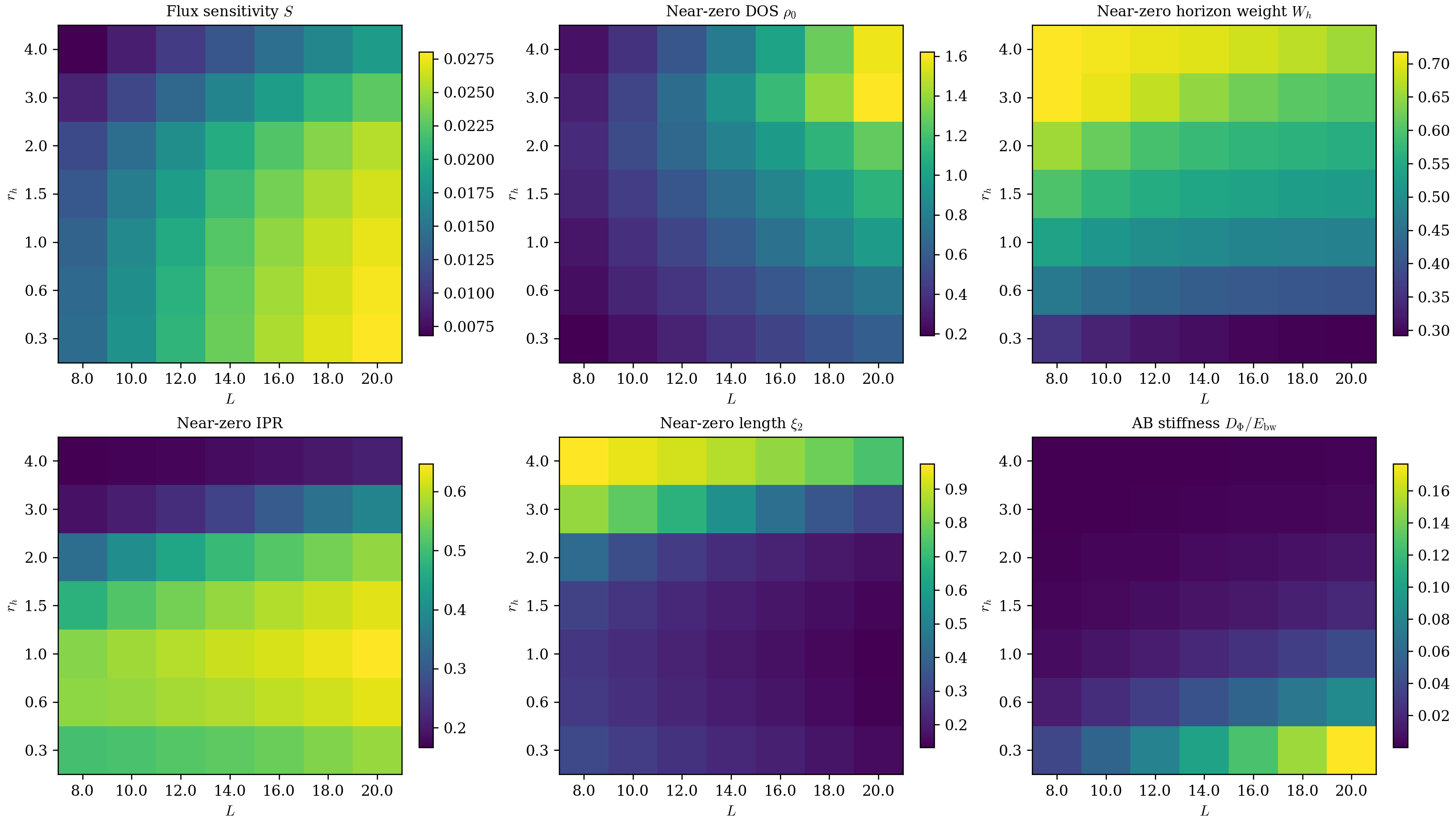}
\caption{Full-exterior diagnostics including the horizon ($x_0=0$): the global flux sensitivity $S$, the near-zero DOS $\rho_0$, the near-zero horizon weight $W_h$, the near-zero IPR, the second-moment length $\xi_2$, and the Aharonov--Bohm stiffness magnitude $\mathcal D_{\Phi}$. Larger $r_h$ increases the horizon-localized fraction of the near-zero sector while suppressing both magnetic and cycle response.}
\label{fig:fullmetrics-strong}
\end{figure}

The global parameter scans are summarized in Figs.~\ref{fig:annulusmetrics-strong} and \ref{fig:fullmetrics-strong}. On the annulus, larger $L$ tends to sharpen the fragmented spectrum: the box dimension $D_{\rm box}$, the global flux sensitivity $S$, the near-zero DOS $\rho_0$, and the Aharonov--Bohm stiffness $\mathcal D_{\Phi}$ all increase as the local curvature weakens. This is the global version of what is already visible in the color-coded spectra. In effect, the annular heatmaps coarse-grain the same dewarping trend already visible by eye: when $L$ grows, the BTZ coefficients vary more slowly with radius and the butterfly moves closer to the less distorted flat-limit pattern.

On the full exterior the trend with $r_h$ is different. The horizon-weight heatmap shows that increasing $r_h$ pushes more of the near-zero sector into the near-horizon region. At the same time both the magnetic sensitivity $S$ and the cycle response $\mathcal D_{\Phi}$ decrease. The full-exterior scan therefore supports a simple interpretation: $L$ controls how sharply the butterfly fragments, while $r_h$ controls how much of the low-energy sector becomes redshift-dominated and weakly dispersing. The key point is that these are not separate empirical tendencies. Larger $r_h$ simultaneously increases $W_h$ and $\rho_0$ because more states accumulate near $E\approx 0$ in the throat, and it decreases $S$ and $\mathcal D_{\Phi}$ because those same states sit where the angular scale $t_{\phi}(x)$ is smallest.

Finally, Fig.~\ref{fig:robust-strong} addresses the main numerical caveats. The qualitative conclusions remain stable under moderate changes in lattice size and flux-grid resolution. We also compare different $x_0$ values at fixed $x_{\max}$, rather than at fixed $(N_x,a_x)$. This removes the most obvious finite-window ambiguity of the earlier horizon comparison and shows that the main trend---decreasing response along with increasing horizon localization---survives the fixed-window check as well.

\begin{figure}[t]
\centering
\includegraphics[width=0.96\textwidth]{btz_robustness.png}
\caption{Robustness checks at a representative full-exterior point. Left: convergence with lattice size at fixed radial window. Middle: convergence with the number of sampled $\alpha_B$ values. Right: comparison of different $x_0$ values at fixed $x_{\max}$. The plotted quantities are normalized to the first point in each series.}
\label{fig:robust-strong}
\end{figure}

Overall, the numerical results support the following picture. The AdS radius $L$ controls the local curvature and therefore how strongly the BTZ butterfly is warped away from the flat Harper limit. The horizon radius $r_h$ controls the throat size and the near-horizon redshift, and therefore how much of the low-energy sector becomes weakly sensitive to both magnetic flux and Aharonov--Bohm twist. The resulting curved Harper problem is thus naturally associated with BTZ geometry, but it should be interpreted as a BTZ-derived effective single-band model rather than as the direct spectrum of the continuum BTZ Dirac Hamiltonian.

\section{Conclusion}\label{sec:conclusion}

In this paper we derived the reduced Dirac Hamiltonian on the non-rotating BTZ background and used its redshift structure to construct a gauge-covariant effective lattice model on the constant-time BTZ cylinder. In equal-area coordinates the geometric meaning of the two main parameters is especially transparent: the AdS radius $L$ fixes the local Gaussian curvature, while the horizon radius $r_h$ fixes the throat size and the strength of the near-horizon redshift. The angular Fourier transform of the lattice problem then yields an exact curved Harper equation with BTZ-dependent hopping amplitudes, a consistent dimensionless angular quasi-momentum, and a natural distinction between Ramond and Neveu--Schwarz sectors. The resulting construction should be viewed as a BTZ-derived effective single-band model whose relation to geometry is explicit at every stage.

The numerical analysis shows that the two geometric scales control different aspects of the spectrum. Increasing $L$ weakens the local curvature and sharpens the butterfly-like fragmentation toward the flat Harper limit. Increasing $r_h$, by contrast, leaves the local curvature fixed but enlarges the throat and strengthens the redshift effect on low-lying states, producing a larger near-horizon sector that is weakly dispersing and only weakly sensitive to both magnetic flux and Aharonov--Bohm twist. The state-resolved diagnostics make this mechanism visible directly: states with the smallest mean radius carry the largest near-horizon weight and the weakest flux response, while the global scans show the corresponding suppression of both magnetic sensitivity and cycle stiffness. The BTZ geometry thereby transcends a mere deformation of the canonical Hofstadter problem, instead orchestrating a distinctive interplay among negative curvature, horizon localization, spin structure, and cycle response.

More broadly, this work provides a concrete interface between AdS black-hole geometry and hyperbolic band theory. It suggests several natural extensions: rotating or charged black-hole backgrounds, a direct discretization of the full two-component Dirac problem, topological and transport diagnostics in curved magnetic bands, and synthetic realizations in hyperbolic circuit, topolectrical, or quantum-simulation platforms. It would also be interesting to understand how the present effective description connects to broader holographic and quantum-information questions, where redshift, nontrivial cycles, and negatively curved spatial organization already play a central role. Our main message is that BTZ offers a clean laboratory in which ideas from string theory, holography, condensed matter, quantum information, and mathematical physics can be translated into a common spectral language.
Rather than merely deforming a Hofstadter problem, the BTZ geometry engenders a horizon-controlled spectral sector whose dynamical behavior is qualitatively distinct.

\section*{Data and code availability}
Source data and code for the work are available from authors' GitHub repository:\\
\url{https://github.com/IKEDAKAZUKI/Hofstadter-butterfly-in-AdS3}

\section*{Acknowledgments}
This work was supported by the NSF under Grant No. OSI-2328774 (KI), by the Israeli Science Foundation Excellence Center, the US-Israel Binational Science Foundation, the Israel Ministry of Science (YO).

\appendix
\section{Conventions}

Flat-frame gamma matrices satisfy
\begin{equation}
\{\gamma^{a},\gamma^{b}\}= - 2\,\eta^{ab}.
\end{equation}
A convenient \(2\times 2\) representation (Pauli basis) is
\begin{equation}
\gamma^{0}= \sigma_z,\qquad
\gamma^{1}= i\sigma_y,\qquad
\gamma^{2}= i\sigma_x,
\label{gamma}
\end{equation}
which yields \((\gamma^{0})^2=\mathbb{I}\) and \((\gamma^{1})^2=(\gamma^{2})^2=-\mathbb{I}\).

With our hermiticity conditions of the gamma matrices (\ref{gamma}), the hermitian Lorentz generators are:
\begin{equation}
\Sigma^{12} = \frac{i}{4}[\gamma^1,\gamma^2],\qquad
\Sigma^{01} = \frac{1}{4}[\gamma^0,\gamma^1],\qquad
\Sigma^{02} = \frac{1}{4}[\gamma^0,\gamma^2] \ ,
\end{equation}
where $\Sigma^{12}$ is the rotation in the $1-2$ plane, and $\Sigma^{01},\Sigma^{02}$ are the boosts.

The curved spacetime gamma matrices are given by:
\begin{equation}
\gamma^\mu(x)= e_{a}{}^{\mu}(x)\,\gamma^{a},\qquad
\{\gamma^\mu(x),\gamma^\nu(x)\}= -2\,g^{\mu\nu}(x) \ .
\end{equation}
They read:
\begin{equation}
\gamma^t=\frac{1}{\sqrt{f(r)}}\,\gamma^{0}= \frac{1}{\sqrt{f(r)}}\,\sigma_z,\qquad
\gamma^r=\sqrt{f(r)}\,\gamma^{1}= i \sqrt{f(r)}\,\sigma_y,\qquad
\gamma^\phi=\frac{1}{r}\,\gamma^{2}= i \frac{1}{r}\,\sigma_x \ .
\end{equation}

\bibliographystyle{JHEP}
\bibliography{ref}
\end{document}